\documentclass[a4paper,11pt]{article}
\pdfoutput=1 

\usepackage[utf8]{inputenc}
\usepackage{jinstpub} 
\usepackage{subcaption}
\usepackage[table,xcdraw]{xcolor}
\usepackage{lineno}
\usepackage{multirow}

\title{Measurement of the distribution of $^{207}$Bi depositions on calibration sources for SuperNEMO}

\author[a]{R.~Arnold,}
\author[b]{C.~Augier,}
\author[c]{A.S.~Barabash,}
\author[d]{A.~Basharina-Freshville,}
\author[e]{E.~Birdsall,}
\author[b]{S.~Blondel,}
\author[b]{M.~Bongrand,}
\author[b]{D.~Boursette,}
\author[f]{R.~Breier,}
\author[g,h,1]{V.~Brudanin,\note{Deceased.}}
% ...

\affiliation[a]{IPHC, ULP, CNRS/IN2P3, F-67037 Strasbourg, France}
\affiliation[b]{Universit\'{e} Paris-Saclay, CNRS/IN2P3, IJCLab, 91405 Orsay, France}
\affiliation[c]{NRC ``Kurchatov Institute'', ITEP, 117218 Moscow, Russia}
\affiliation[d]{UCL, London WC1E 6BT, United Kingdom}
\affiliation[e]{University of Manchester, Manchester M13 9PL,~United Kingdom}
\affiliation[f]{FMFI,~Comenius~University,~SK-842~48~Bratislava,~Slovakia}
\affiliation[g]{JINR, 141980 Dubna, Russia}
\affiliation[h]{National Research Nuclear University MEPhI, 115409 Moscow, Russia}
% ...

\collaboration{SuperNEMO Collaboration:}

\emailAdd{miroslav.macko@utef.cvut.cz}

\abstract{The SuperNEMO experiment will search for neutrinoless double-beta decay ($0\nu\beta\beta$), and study the Standard-Model double-beta decay process ($2\nu\beta\beta$). 
The SuperNEMO technology can measure the energy of each of the electrons produced in a double-beta ($\beta\beta$) decay, and can reconstruct the topology of their individual tracks.
The study of the double-beta decay spectrum requires very accurate energy calibration to be carried out periodically. The SuperNEMO Demonstrator Module will be calibrated using 42 calibration sources, each consisting of a droplet of $^{207}$Bi within a frame assembly. 

The quality of these sources, which depends upon the entire $^{207}$Bi droplet being contained within the frame, is key for correctly calibrating SuperNEMO's energy response.
In this paper, we present a novel method for precisely measuring the exact geometry of the deposition of $^{207}$Bi droplets within the frames, using Timepix pixel detectors. We studied 49 different sources and selected 42 high-quality sources with the most central source positioning.
}


\begin{document}
\maketitle
\flushbottom

\section{Introduction}

\subsection{The SuperNEMO experiment} 

SuperNEMO \cite{SUPERNEMO, SNEMO-DESCR} is a double-beta ($\beta\beta$) decay experiment, designed to look for the hypothesized lepton-number-violating process of neutrinoless double-beta decay ($0\nu\beta\beta$). SuperNEMO's tracker-calorimeter design, based on the NEMO-3 technology, is also well suited for precision studies of the Standard Model double-beta decay process ($2\nu\beta\beta$) which is present in all $0\nu\beta\beta$ candidate isotopes. Both types of $\beta\beta$ decay produce two electrons around the MeV energy scale, which SuperNEMO can individually track; the processes can be distinguished by studying the energies of these electrons.

\begin{figure}
            \centering
  			\includegraphics[width=0.72\textwidth]{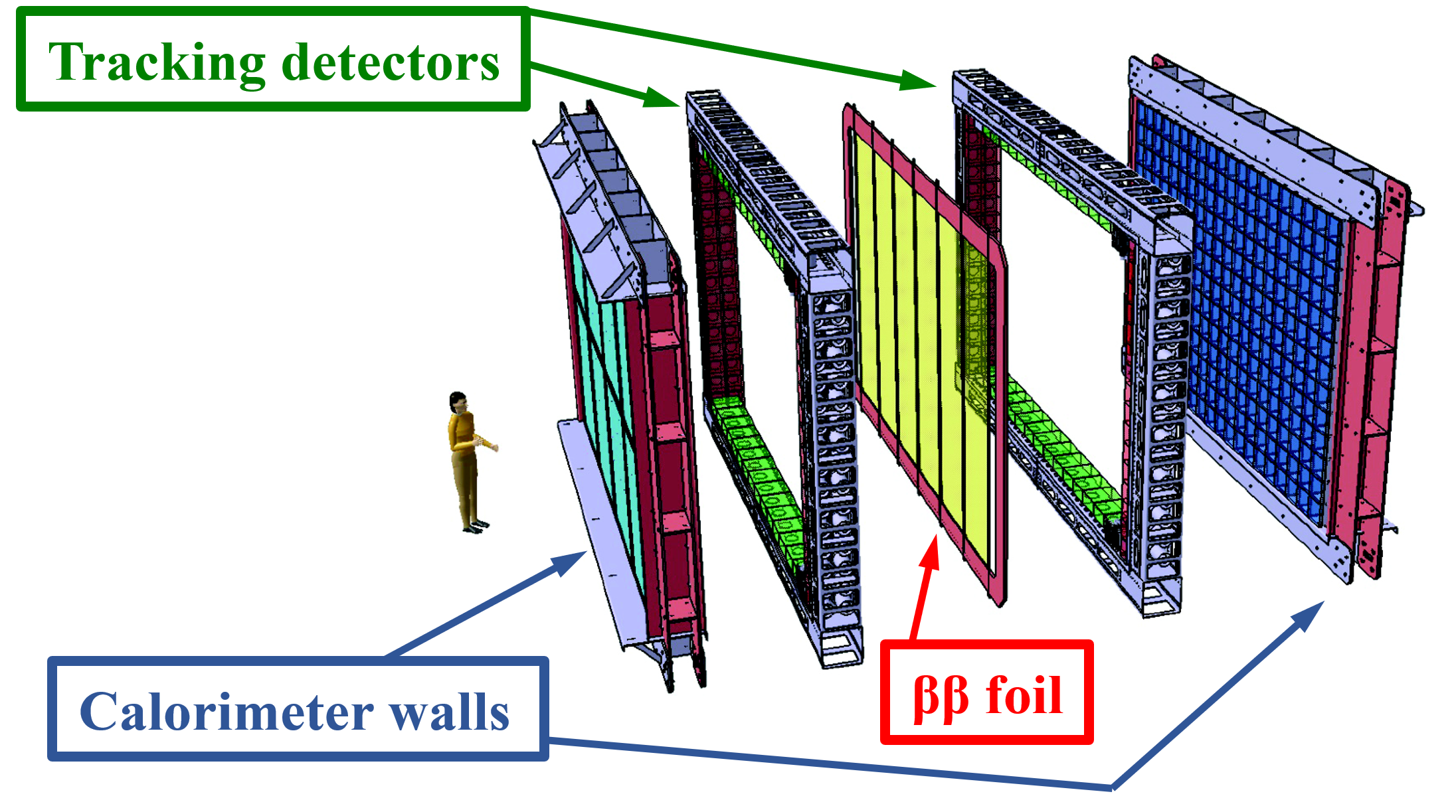}
            \caption{Overview of the SuperNEMO Demonstrator Module. Figure by C. Bourgeois \cite{CALORIMETER}.}
            \label{fig:SNmodule}
        \end{figure}

The majority of $0\nu\beta\beta$ detectors are homogeneous, meaning that the  $\beta\beta$ source also serves as the detection material (for example $^{76}$Ge semiconductor detectors, or bolometers of $\beta\beta$ isotope-enriched crystals).
In SuperNEMO, however, the source is independent of the detector. The basic unit of a SuperNEMO-style detector is a module, as shown in figure \ref{fig:SNmodule}. The  modular design allows the detector size to be increased by  adding identical modules to the detector. After the construction of each module, it is possible to verify whether the expected specifications (e.g. energy resolution, background level, etc.) have been met. An initial module, known as the SuperNEMO Demonstrator, is currently undergoing the final stages of installation and commissioning at LSM (the Modane Underground Laboratory, France). 

A SuperNEMO module  consists of a $\beta\beta$-decay source in the form of thin, solid foils enriched in a $\beta\beta$-decaying isotope, sandwiched between two identical detector halves. This opens up the possibility of studying any $\beta\beta$-decaying candidate isotope that can be produced in the form of thin foils. The SuperNEMO Demonstrator Module has 34 source foils amounting to a total of 6.25\,kg of selenium, of which 6.11\,kg is the $\beta\beta$ isotope $^{82}$Se, with an average thickness of 0.286\,mm and an average surface density of 50\,mg of selenium per cm$^2$ \cite{FOILS, SOURCE_PROD}. The feasibility of using other isotopes (such as $^{150}$Nd) is also under investigation.

The two electrons emitted in $\beta\beta$ decays are detected in the SuperNEMO Demonstrator Module by its particle tracker and calorimeter. The source foils are sandwiched between two tracking detectors, which together consist of 2034 drift cells operating in Geiger mode, in a vertical magnetic field \cite{TRACKER}. The tracker reconstructs charged-particle trajectories, allowing significant background suppression, as well as extracting the angular distribution of the $\beta\beta$ electrons. 
 The energy of the individual decay electrons is measured by a segmented calorimeter, consisting of plastic scintillator blocks coupled to photomultiplier tubes surrounding the detector on all six sides. Calorimeter modules in the main walls of the detector have an average energy resolution of $7.5\%$\,FWHM at 1\,MeV. The ability of the calorimeter to distinguish a $0\nu\beta\beta$ signal from $2\nu\beta\beta$, and to  study $2\nu\beta\beta$ decay mechanisms that affect the energy spectrum, depends on the correct calibration of its energy response \cite{CALORIMETER}.

\subsection{SuperNEMO calibration system} 

\begin{figure}
            \centering
            \begin{subfigure}[b]{0.3\textwidth}
                    \centering
  				    \includegraphics[width=\textwidth]{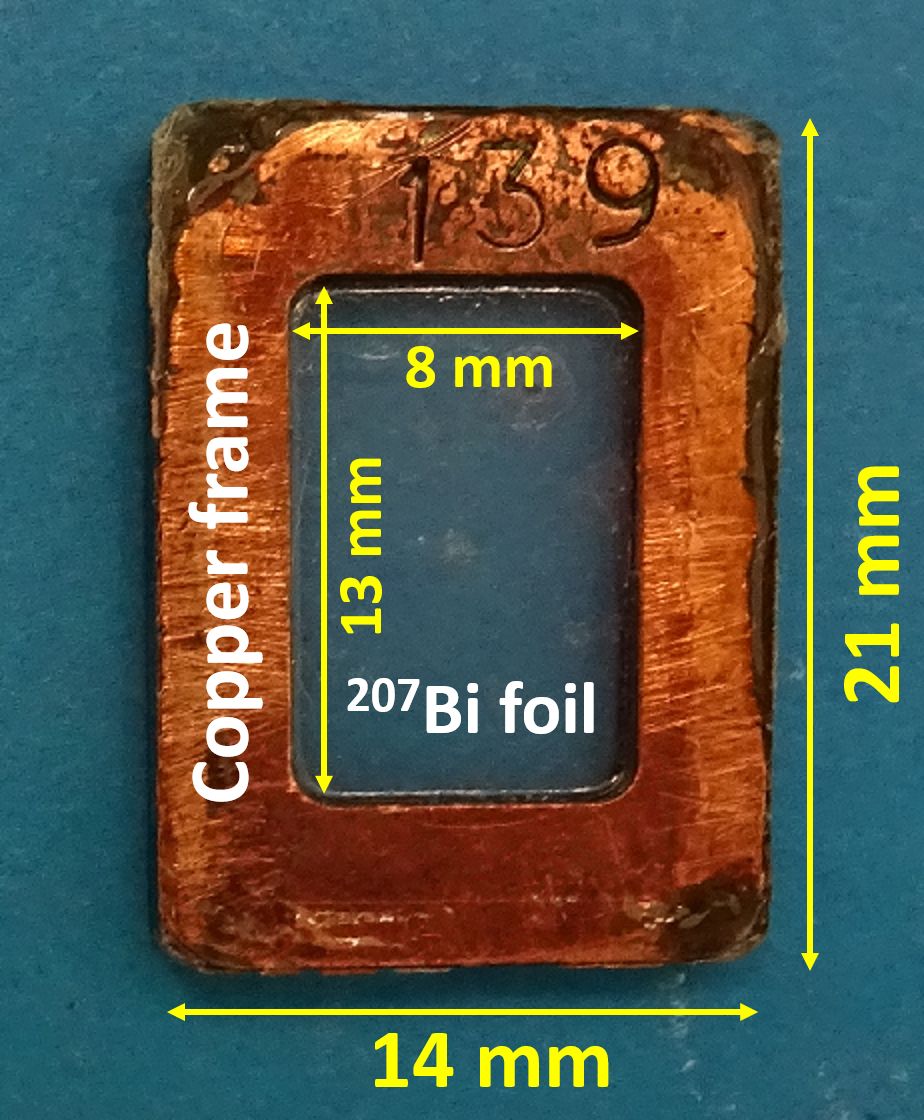}
                    \caption{}
                    \label{fig:syst_A}
			\end{subfigure}
			\begin{subfigure}[b]{0.17\textwidth}
			        \centering
  				  \includegraphics[width=\textwidth]{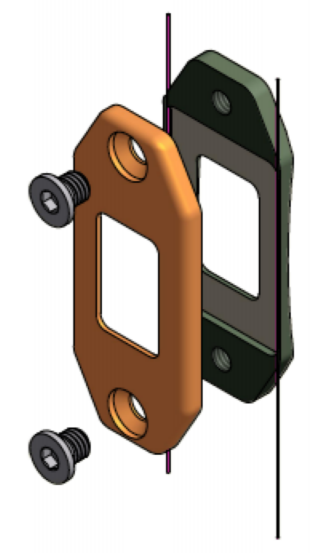}
                    \caption{}
                    \label{fig:syst_B}
			\end{subfigure}
			\begin{subfigure}[b]{0.5\textwidth}
			        \centering
  			         \includegraphics[width=\textwidth]{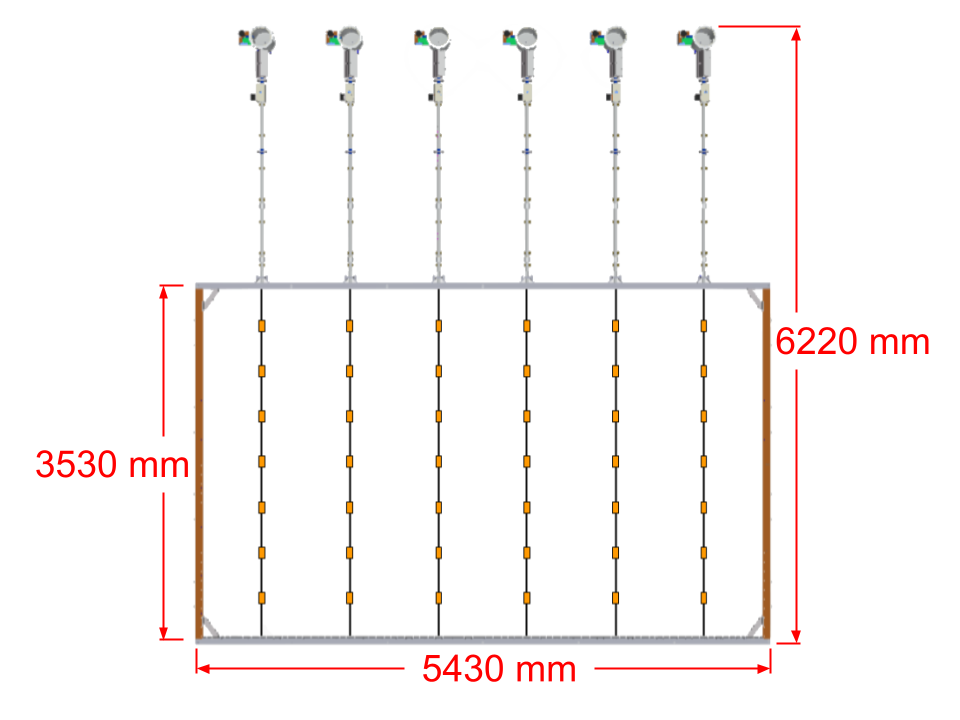}
                    \caption{}
                    \label{fig:syst_C}
			\end{subfigure}
            \caption{a) Photograph of $^{207}$Bi source number\,139. b) Diagram of the envelope which fixes the source within the deployment system. c) Simplified diagram of the $^{207}$Bi deployment system. Orange rectangles represent source envelopes (not to scale). Six columns, each comprising seven  $^{207}$Bi sources, are vertically deployed into gaps between the  $^{82}$Se $\beta\beta$ source foils by an automatic system above the SuperNEMO Demonstrator Module.}
            \label{fig:syst}
		\end{figure}

In order to perform energy calibrations of the SuperNEMO Demonstrator Module, $^{207}$Bi sources are used. K-shell electrons from internal-conversion decays of $^{207}$Bi provide three relevant calibration lines at 482\,keV, 976\,keV and 1682\,keV; lower-intensity lines from L- and M-shell conversions are also included to achieve a good fit of the  $^{207}$Bi spectrum. The sources, which were previously used in the NEMO-3 detector\cite{NEMO3}, were produced at IPHC Strasbourg. For each source, a very thin droplet of $^{207}$Bi solution was deposited on a Mylar foil (12\,$\mu$m thick) and then covered with another identical foil. Over time, the solvent has dried out, but the droplets of $^{207}$Bi remain stable. Both Mylar foils were sealed by a rectangular frame made of radiopure copper, with internal dimensions 8\,mm\,$\times$\,13\,mm  (figures \ref{fig:syst_A} and \ref{fig:syst_B}). The dimensions of the sources were chosen for compatibility with NEMO-3's calibration tubes, and have been incorporated into the SuperNEMO design.

The 42 calibration sources used in the SuperNEMO Demonstrator Module are inserted and guided into gaps between the $^{82}$Se $\beta\beta$ source foils, using six identical mechanisms (figure \ref{fig:syst_C}). The insertion and removal of the sources will be controlled by an automatic source deployment system developed by the University of Texas at Austin \cite{DEPLSYST}. The calibration procedure will be performed at regular intervals during the data taking; the necessary calibration frequency will be determined experimentally, and is expected to be on the order of one run every few days or weeks. Information about the position of the sources relative to the rest of the detector is used when reconstructing the emitted electron trajectories and the associated energy losses. The deployment system lowers the source envelopes to their predefined locations as shown in figure \ref{fig:syst_C}, with a precision of 60\,$\mu$m. However, the deposition distribution of the $^{207}$Bi droplet within the copper frame is unknown (figure \ref{fig:syst_A}). Most importantly, one needs to verify that the $^{207}$Bi droplet did not spill in between the copper frame parts during the sealing process. This could cause emitted electrons to lose additional energy, which would result in a distortion of the reconstructed energy lines, and spoil the energy calibration of the  SuperNEMO Demonstrator Module.
        
The goal of this work was to establish an experimental method to describe the deposition distribution of $^{207}$Bi droplets within the calibration sources for the SuperNEMO Demonstrator Module.  $^{207}$Bi is a complex emitter; along with electrons, it emits both X-rays and gammas. 
For this analysis, instead of observing the electrons used to calibrate the SuperNEMO Demonstrator Module, the $^{207}$Bi deposition within each source was studied by detecting low-energy X-rays from the $^{207}$Bi source, which produce single-pixel depositions in a silicon pixel detector, allowing us to take advantage of the detector's peak sensitivity, and providing better localisation of signals. As a result, we established a ranking of the measured SuperNEMO calibration sources based on how closely the $^{207}$Bi droplet had been deposited relative to the center of the copper frame. The best 42 sources were chosen to be  installed in the calibration setup for the SuperNEMO Demonstrator Module. 

\section{Measurement and data analysis}
\subsection{Timepix Detectors and Calibrations}
\label{subsec:calib}
\begin{figure}
            \centering
            \begin{subfigure}{0.25\textwidth}
                \centering
  			    \includegraphics[width=\textwidth]{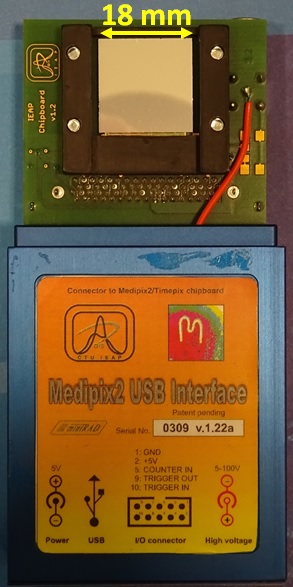}
                \caption{}
			\end{subfigure}
  			\begin{subfigure}{0.485\textwidth}
  			    \centering
  			    \includegraphics[width=\textwidth]{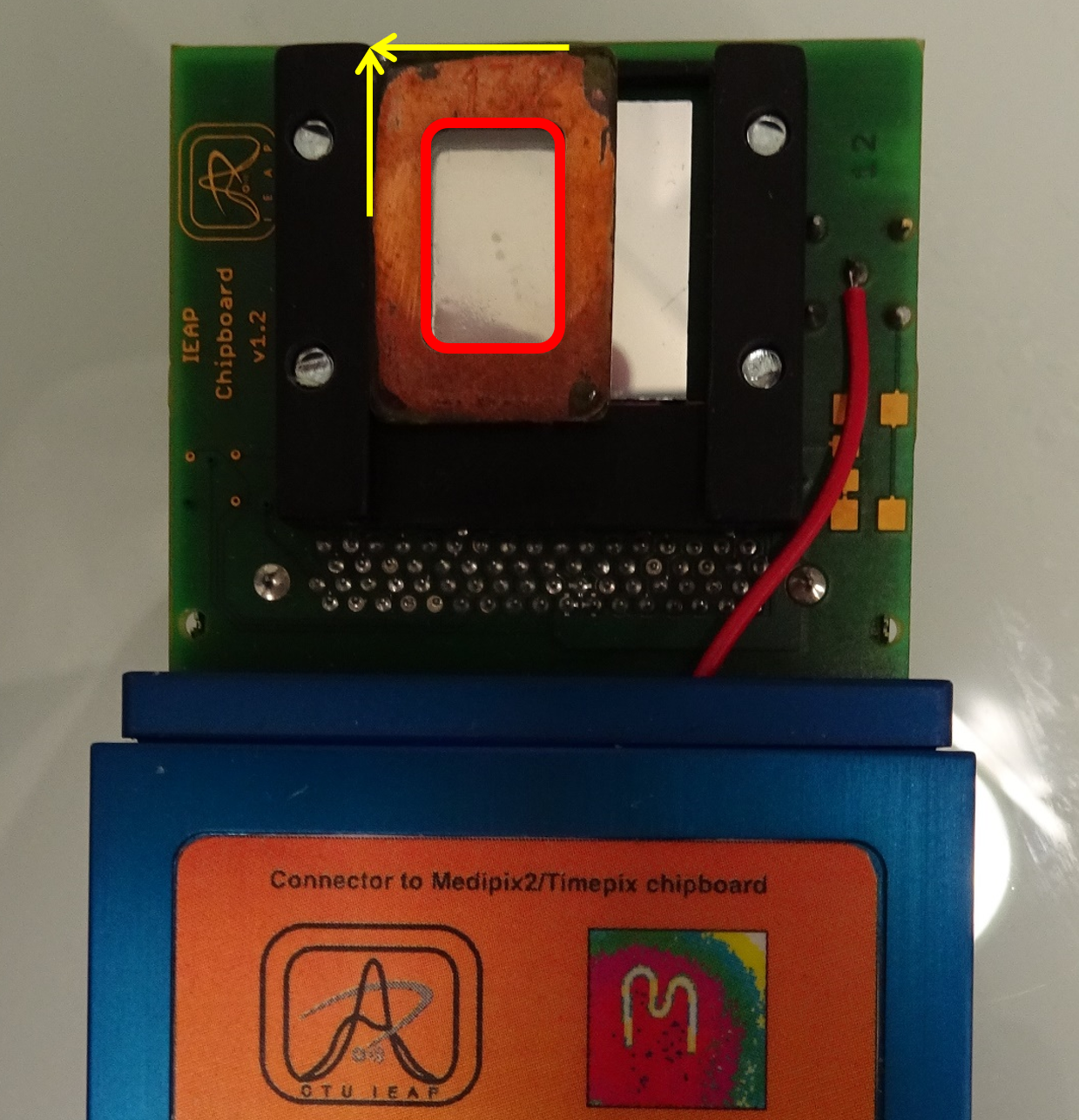}
                \caption{}
                \label{fig:setup_B}
            \end{subfigure}
            \caption{a) Photograph of one of the Timepix pixel detectors used in this work, H04-W0163, which has a 0.3\,mm-thick silicon sensor. 
            b) Photograph of the source position during each measurement. The two yellow arrows point to the Reference Alignment Point (RAP).}
            \label{fig:det}
        \end{figure}

The droplet positioning measurements were performed using three Timepix silicon pixel detectors provided by IEAP, CTU in Prague (figure \ref{fig:det}). Two of the detectors (H04-W0163 and H11-W0163) are 0.3\,mm thick, while the third (L05-W0163) is 1\,mm thick. Timepix detectors have the ability to measure the energy deposited by an incident particle, and its time of arrival. They were developed by the Medipix collaboration \cite{TIMEPIX}. The sensitive chip is a square with sides 14.08\,mm long, divided into 256~$\times$~256 pixels (size of each pixel: 0.055\,mm~$\times$~0.055\,mm).

For the Timepix detectors to be used to measure the distribution of the $^{207}$Bi source droplets, the detectors required a precise spatial calibration, and an approximate energy calibration. Both were carried out at IEAP, CTU in Prague. The energy response of each of the 65536~pixels in each detector was calibrated individually. This calibration consisted of relatively long measurements (several hours) in order to collect enough statistics in each pixel. The calibration was performed using fluorescent foils exposed to X-rays. Three fluorescent foils --- Fe (6.398\,keV), Cu (8.04\,keV) and Cd (23.106\,keV) --- each provided one energy calibration point. The calibration function has four parameters; the fourth parameter is the result of threshold equalization of the detector. \cite{CALIB_FUNCT}. This energy calibration  allowed the energy-weighted distribution of spatial data from the $^{207}$Bi sources to be generated, and enabled a basic energy cut to select the particles of interest for our $^{207}$Bi measurement. By using low-energy X-rays, we were able to select single-pixel depositions, which give us the best position localization. Higher-energy depositions, like those from $\beta$-decay electrons, activate several pixels, which reduces the ability to identify the spatial coordinates of the decay.

When a Timepix detector is used to monitor the $^{207}$Bi source, the $^{207}$Bi droplet position is only represented in the detector chip coordinates. However, the purpose of the measurement is to represent the position of the $^{207}$Bi droplet with respect to its copper frame. In order to make such spatial calibration possible, we defined a Reference Alignment Point (RAP, figure \ref{fig:setup_B}) for all measurements. 

\begin{figure}
            \centering
            \begin{subfigure}[b]{0.48\textwidth}
                \centering
  			    \includegraphics[width=\textwidth]{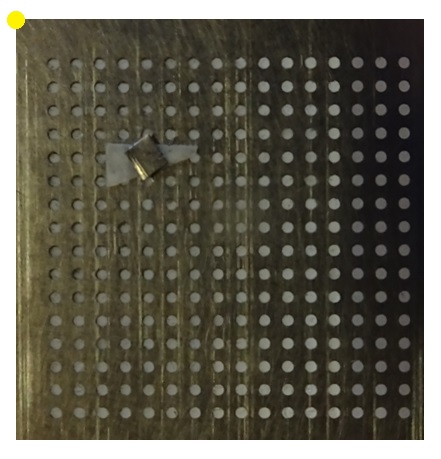}
                \caption{}
                \label{fig:grid_A}
			\end{subfigure}
			\begin{subfigure}[b]{0.445\textwidth}
			    \centering
  			    \includegraphics[width=\textwidth]{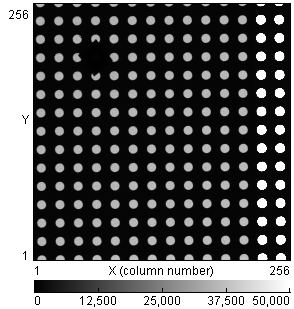}
                \caption{}
                \label{fig:grid_B}
			\end{subfigure}
            \caption{a) Photograph of the calibration grid. The yellow circle denotes the corner corresponding to the RAP during the spatial calibration measurements. b) The grid as seen in the detector after the exposure to X-rays. Neither the first two rows nor the first column are visible in the dataset; nor is the yellow reference point shown in the left-hand figure, as these lie outside the sensitive area of the detector. In order to identify individual holes in the dataset, we covered one of the holes with a piece of metal.}
            \label{fig:grid}
        \end{figure}

An additional measurement was taken in which a thin metallic square, perforated by holes distributed equidistantly in a  grid (figure \ref{fig:grid_A}) laser-cut to few-$\mu$m precision, was placed on top of the detector, so that the yellow dot in figure \ref{fig:grid_A} matched the RAP. The distance between the centers of adjacent holes, in both the vertical and horizontal directions, was equal to 1\, mm. We irradiated this setup with X-rays. The result of this measurement can be seen in  figure \ref{fig:grid_B}. Thanks to this measurement, we were able to extract the position of each pixel in the sensitive area of the detector relative to the RAP. The position of the RAP and the 1-mm spacing of the metal grid were used to convert from the coordinate system of the detector (in pixels) to that of the source frame (in mm).

\subsection{Measurement protocol}

The  duration of a measurement sufficient to collect enough data for an individual $^{207}$Bi source  was typically around  2--3 hours, due to the relatively low source activity (120-145\,Bq) and selection efficiency (approximately 6\% for the thinner detectors, and 9\% for the thicker, for the data of interest to this study: while the detector's efficiency is close to 100\% for X-rays produced by $^{207}$Bi, these values refer to the efficiency of the single-pixel and energy-range cuts imposed to ensure that the position of the droplet corresponds to the activated pixel). This led to an event rate in our region of interest (ROI), as explained in section \ref{subs:datana}, of around 8--10 events per second.  To accumulate sufficient statistics to precisely measure the droplet position, we aimed for around 50000 events in the ROI for each source. Due to scheduling constraints, less data was collected for sources 93, 94, and 95; nevertheless, it was sufficient to evaluate these sources. Repeated and extended measurements were taken for a subset of sources, taking advantage of the laboratory schedule to collect larger samples with millions of events and confirm consistency with the lower-statistics measurements.
 All three pixel detectors were used in parallel, and were connected to one computer. Measurements for the 49 available $^{207}$Bi sources were taken over a period of almost 12 full days, which included short interruptions to exchange the samples. 

\subsection{Structure of measured data}

\begin{figure}
		    \centering
  			\includegraphics[width=0.9\textwidth]{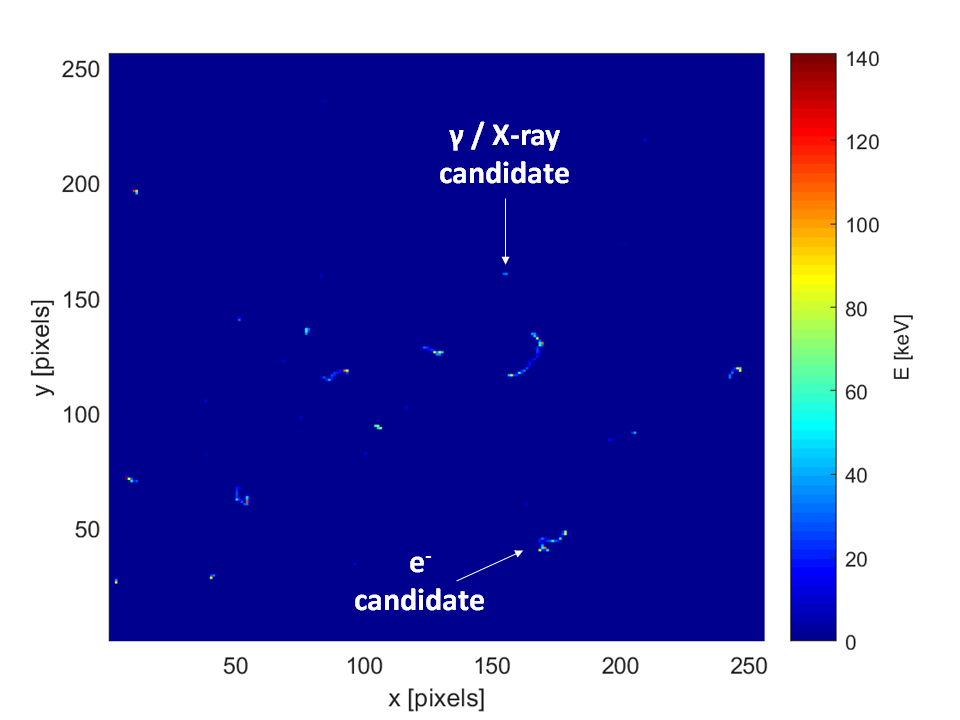}
            \caption{Visualization of the data collected from a sample $^{207}$Bi source in a one-second time slice. A typical photon and electron candidate are marked.}
            \label{fig:frame}
        \end{figure}

As mentioned above, each of the Timepix pixel detectors used in the study has a square chip with a size of 14.08\,mm $\times$ \,14.08\,mm divided into $256\,\times\,256$ pixels.  Raw data obtained from a detector has a simple format of three columns, storing the $x$ and $y$ coordinates of each pixel and the energy deposited in the pixel, respectively. Each file contains information about the total energy per individual pixel over a specified time period known as a ``time slice''. The dead time between two time slices is typically around 200\,ms, regardless of the length of the time slice. A time slice of 1\,second was chosen, as this allowed individual particles to be isolated, while minimizing the relative fraction of dead time. In figure \ref{fig:frame}, data recorded in a single 1-second time slice is visualized as an example. 

The software used, Pixelman \cite{PIXELMAN} (developed at IEAP, CTU in Prague), provides a user interface for the Timepix detectors. While the software can be used to extract the previously-described raw data, Pixelman also provides a pre-analyzed `clustered output' \cite{CLUST} format, which enables the identification and energy measurement of individual particles. The algorithm that transforms the raw data into clustered output first groups the triggered pixels from one time slice into clusters, based on their spatial adjacency. It then extracts the energy of each cluster (representing one particle, which might have triggered several pixels) and its position. The sum of the energy deposited in all the individual pixels of a cluster represents the energy deposited by the particle in the detector. The cluster position is defined as the average position of the individual pixels in the cluster, weighted by the energy deposited in each pixel. The clustered output also reports a `cluster size', corresponding to the number of pixels in each cluster, which is related to the particle type: single-pixel isolated energy deposits are typically produced by photons and Auger electrons, while high-energy electrons generate multi-pixel clusters.   The analysis described in the following sections was performed on this clustered output.
        
\subsection{Data analysis method}
\label{subs:datana}
The main goal of this study was to develop an experimental method that could evaluate the quality of a source based on the position of $^{207}$Bi within its source frame assembly. In practice, this can prevent the use of any sources with unwanted contamination caused by $^{207}$Bi droplet leakage into the region of the copper source frame. The study also has the potential to reduce systematic uncertainties related to the calibration source positions within the SuperNEMO Demonstrator Module. Precise knowledge of source positions can enable us to evaluate our tracker's vertex resolution capabilities, and to ensure that electron track lengths are correctly reconstructed.

\begin{figure}
            \centering
  			\begin{subfigure}[b]{0.48\textwidth}
  			    \centering
  			    \includegraphics[width=\textwidth]{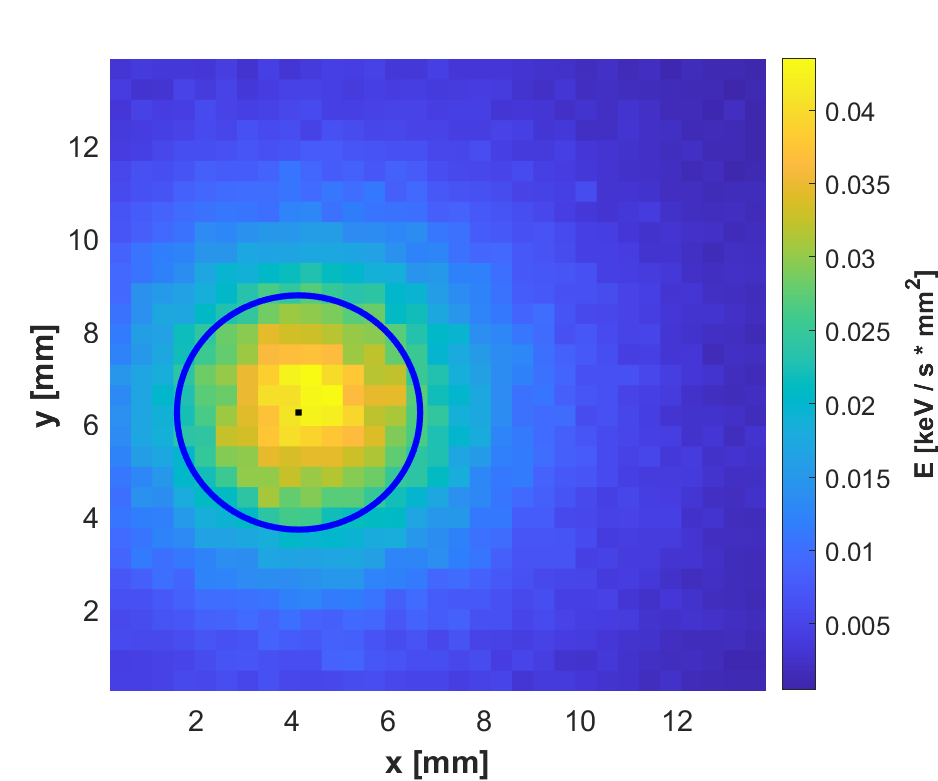}
  			    \caption{}
  			    \label{fig:cuts_A}
			\end{subfigure}
			\hspace{10pt}
			\begin{subfigure}[b]{0.48\textwidth}
			    \centering
  			    \includegraphics[width=\textwidth]{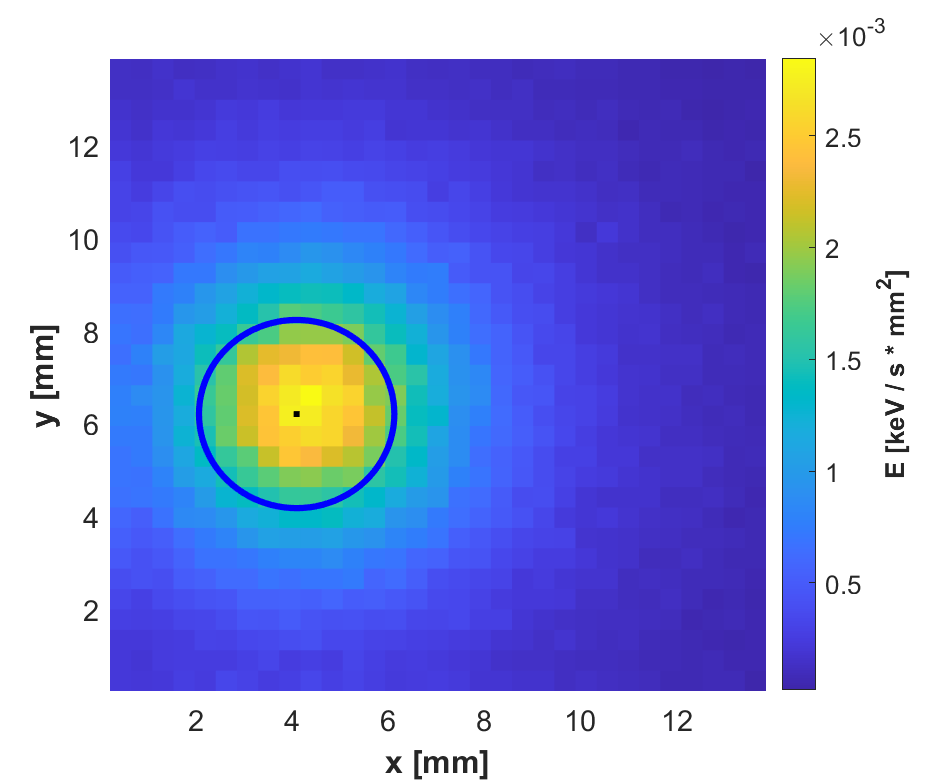}
  			    \caption{}
  			    \label{fig:cuts_B}
			\end{subfigure}
            \caption{Energy distributions measured for source number 133, with different selections applied. The black points represent the position of the calculated weighted center ($x_{0}$, $y_{0}$). The blue circles represent the contour lines where the fitting function reaches half of the maximal value (thus, the radius of the circle is the half width at half maximum, HWHM). a) Energy distribution for all clusters with an energy in the range 3--1300\,keV (no cluster size cut). The distribution includes all types of particle produced in $^{207}$Bi decays (X-rays, gammas and electrons). b) Energy distribution for single-pixel clusters with an energy in the range 3--30\,keV, corresponding to Auger electrons and X-rays. Note differing scale on color axis.}
            \label{fig:cuts}
        \end{figure}

The positions of the clusters measured for a single source (integrated over all the time slices) can be represented by a two-dimensional histogram. Each cluster is weighted by its energy in order to obtain an energy distribution. This histogram shows a wide peak corresponding to the region of the source frame where the $^{207}$Bi has been deposited (figure \ref{fig:cuts}). The $x$ and $y$ directions are aligned with the 8-mm and 13-mm axes of the source frame, respectively.  

\begin{figure}
            \centering
  			\begin{subfigure}{0.48\textwidth}
  			    \centering
  			    \includegraphics[width=\textwidth]{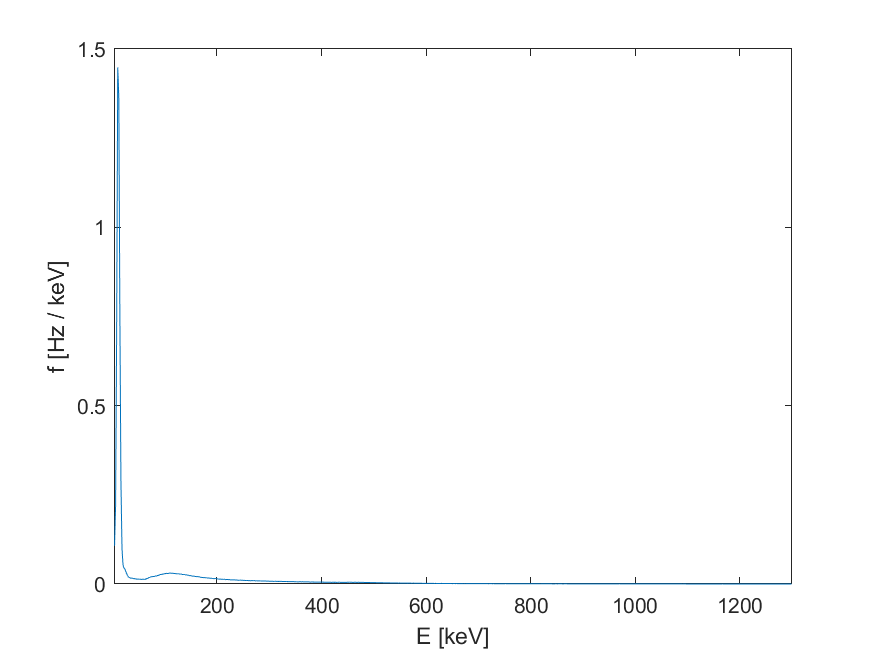}
  			    \caption{}
  			    \label{fig:spectrum_A}
			\end{subfigure}
			\hspace{10pt}
			\begin{subfigure}{0.48\textwidth}
			    \centering
  			    \includegraphics[width=\textwidth]{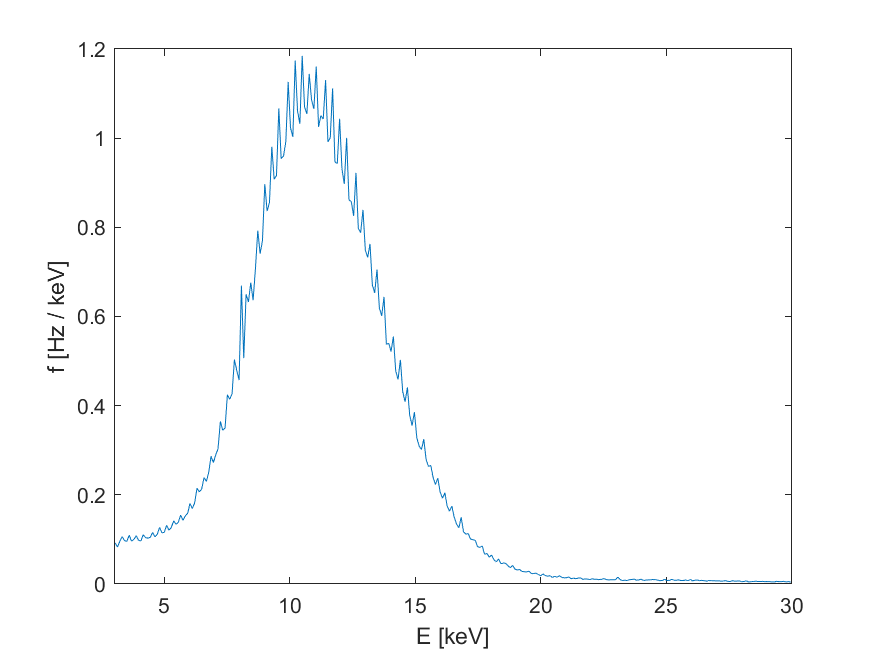}
  			    \caption{}
  			    \label{fig:spectrum_B}
			\end{subfigure}
            \caption{Measured energy spectrum for source number 126, including a) all clusters in the energy range 3--1300\,keV (no cluster size cut) and b) single-pixel clusters in the range  3--30\,keV.}
            \label{fig:spectrum}
        \end{figure}

High-energy electrons activate several pixels; the location of the earliest hit, likely to correspond to the electron's source position in the $^{207}$Bi droplet, is unknown. The best source localization of the $^{207}$Bi droplet was therefore achieved by selecting only the clusters corresponding to a single pixel, and with a reconstructed energy between 3\,keV and 30\,keV, reducing the uncertainty on the measurement of the cluster position (figure \ref{fig:cuts_B}). This energy region consists of X-rays and Auger electrons from electron capture on $^{207}$Bi, and contains the majority of these single-pixel events (figure \ref{fig:spectrum}). This selection made a negligible difference to the best-fitted position coordinates of the droplets' centers, with an average difference between the pre-cut and post-cut position of 0.054\,mm, comparable to the size of a single detector pixel. This energy range and single-pixel cut will henceforth define our region of interest (ROI). It should be noted that, during measurement, there is a small gap (1.5--3\,mm, depending on the individual source) between the $^{207}$Bi source and the Timepix detector. As particles are emitted from the source at varying angles, even when a single pixel is triggered, the corresponding particle may not have been emitted immediately above that pixel.

\begin{figure}[ht]
            \centering
  			\includegraphics[width=0.32\textwidth]{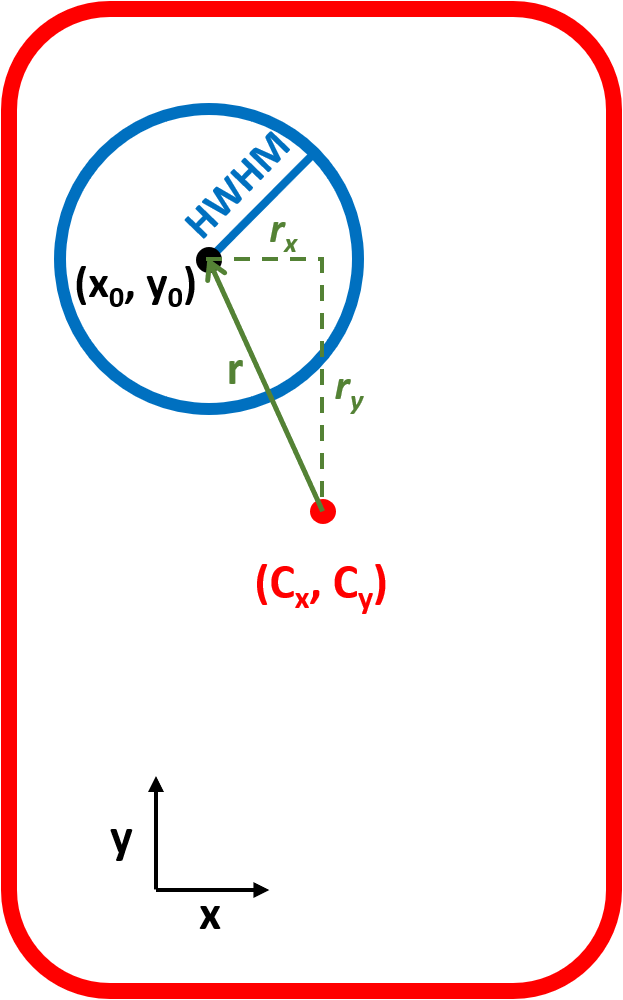}
            \caption{Graphical representation of quantities extracted from the fit to equation \ref{eq:fit}: the blue circle represents the HWHM contour; the black dot denotes the center of the $^{207}$Bi deposition ($x_{0},y_{0}$); the red dot, the center of the frame ($C_{x},C_{y}$); and the green arrow, the  vector $\vec{r}$ of the droplet's position with respect to the frame's center.}
            \label{fig:def}
  \end{figure}

The comparison of energy distributions with and without these cuts is shown in figure \ref{fig:cuts}. After applying the cuts (figure \ref{fig:cuts_B}), the radius of the half-width-at-half-maximum (HWHM) contour is reduced by approximately 20\% with respect to the full sample (figure \ref{fig:cuts_A}). A plot similar to figure \ref{fig:cuts_B} was obtained for each measured source and fitted with a two-dimensional function.
\begin{equation}
	f(x,y)=\frac{A}{\left(x-x_{0}\right)^2+\left(y-y_{0}\right)^2+\gamma} \, .
	\label{eq:fit}
\end{equation}  
Here,the $x_{0}$ and $y_{0}$ coordinates represent the measured center of the source (in pixels), $\gamma$ is proportional to the square of the width of the peak, and $A$ is a scale factor. A visual representation of the quantities extracted from the fit can be found in figure \ref{fig:def}. The blue circles in figures \ref{fig:cuts} and \ref{fig:def} represent the border where the fitting function drops to half of its peak value (i.e. $A/2\gamma$). The radius of the blue circle represents a two-dimensional equivalent of the HWHM  of the distribution. For the purpose of this study, we considered this contour as a measured effective size of the $^{207}$Bi droplet. The fitting function was chosen for its simple form that contains all necessary degrees of freedom. All of the droplets were confirmed visually to be circular in shape, apart from two; both sources with non-circular drops also failed subsequent selection criteria, and were not installed in the SuperNEMO Demonstrator Module.  

The position of the source extracted from the fit is given relative to the sensitive detector chip. Since the aim of the study was to extract the relative position of the source droplet with respect to the source frame, a spatial calibration, described in section \ref{subsec:calib}, was used to provide the relative position of the center of the source frame ($C_{x}$, $C_{y}$) with respect to the chip. Information about the center of the frame allows us to define the position, $\vec{r}$, of the $^{207}$Bi droplet relative to the center of the source frame. It can be calculated as follows:
    \begin{equation}
        \vec{r}=(r_{x}, r_{y})=(x_{0}-C_{x}, y_{0}-C_{y}).
    \end{equation}
The vector $\vec{r}$ can be seen in figure \ref{fig:def}. 

\section{Results}
\label{sec:results}
\subsection{Categorization of sources based on droplet position}
\label{subs:categorize}
For this study, we collected data from 49 different source samples in 52 measurements. 
As explained previously, the standard acquisition time to collect sufficient statistics was 2--3 hours. Supplementary high-statistics measurements on sources 126, 132 and 139, for periods of up to a few days, aimed to improve precision and better estimate systematic uncertainties, as explained in section \ref{subs:syst}. A sample of some of these measurements is shown in table \ref{tab:stats}; a complete list is in tables A1 and A2 of \cite{MACKO_THESIS}.

\begin{table}[ht]
\centering

\caption{Statistics of a sample set of measurements, including standard measurements of 2--3 hours and extended measurements of 2 days or more. The first column shows the ID number of the measured source, with an asterisk (*) denoting measurements repeated under the same experimental conditions. The second and third columns indicate the Timepix detector and its thickness ($d_\text{DET}$). The table also contains the active exposure time (not including dead time) ($t_\text{live}$), total statistics collected ($N_\text{TOT}$), number ($N_\text{ROI}$) and fraction of events in the ROI, and detection rate. The ROI included all single-pixel clusters in the energy range 3--30\,keV. The full data set (TOT) contains all  3--1300\,keV clusters of 1 to 100 pixels. Note the improved detection rate and fraction in the ROI for the thicker (1\,mm) detector.}
\begin{tabular}[c]{|c|c|c|c|cc|c|c|}
\hline
Source & Detector & $d_\text{DET}$ & $t_\text{live}$ & $N_\text{TOT}$ & $N_\text{ROI}$ & $\frac{N_\text{ROI}}{N_\text{TOT}}$ & $\frac{N_\text{ROI}}{t_\text{live}}$\\
ID&&[mm]&[h]&[cnts]&[cnts]&[\%]&[$s^{-1}$]\\ \hline
73 & \footnotesize H04-W0163   & 0.3 & 1.7 & 1.1$\times 10^{5}$ & 5.0$\times 10^{4}$ & 43.3 & 8.3\\
74 & \footnotesize H11-W0163   & 0.3 & 1.8 & 1.2$\times 10^{5}$ & 5.4$\times 10^{4}$ & 43.4 & 8.4\\
111 & \footnotesize H11-W0163  & 0.3 & 3.8 & 2.9$\times 10^{5}$ & 1.3$\times 10^{5}$  & 43.3 & 9.1\\
120 & \footnotesize L05-W0163  & 1.0 & 12.7 & 1.1$\times 10^{6}$  & 5.5$\times 10^{5}$ & 50.4 & 12.1\\
126 & \footnotesize H04-W0163  & 0.3 & 55.5 & 3.4$\times 10^{6}$ & 1.5$\times 10^{6}$ & 43.0 & 7.4\\
126* & \footnotesize H04-W0163 & 0.3 & 46.2 & 3.0$\times 10^{6}$ & 1.3$\times 10^{6}$ & 42.5 & 7.8\\
132 & \footnotesize H11-W0163  & 0.3 & 54.8 & 4.0$\times 10^{6}$ & 1.7$\times 10^{6}$ & 43.5 & 8.7\\
132* & \footnotesize H11-W0163 & 0.3 & 46.0 & 3.3$\times 10^{6}$ & 1.4$\times 10^{6}$ & 43.3 & 8.7\\
139 & \footnotesize L05-W0163  & 1.0 & 54.8 & 4.7$\times 10^{6}$ & 2.3$\times 10^{6}$ & 49.6 & 11.9\\
139* & \footnotesize L05-W0163 & 1.0 & 46.0 & 4.0$\times 10^{6}$ & 2.0$\times 10^{6}$& 49.5 & 11.9\\ \hline
\end{tabular}
\label{tab:stats}
\end{table}

\begin{figure}[ht]
            \centering
  			\includegraphics[width=\textwidth]{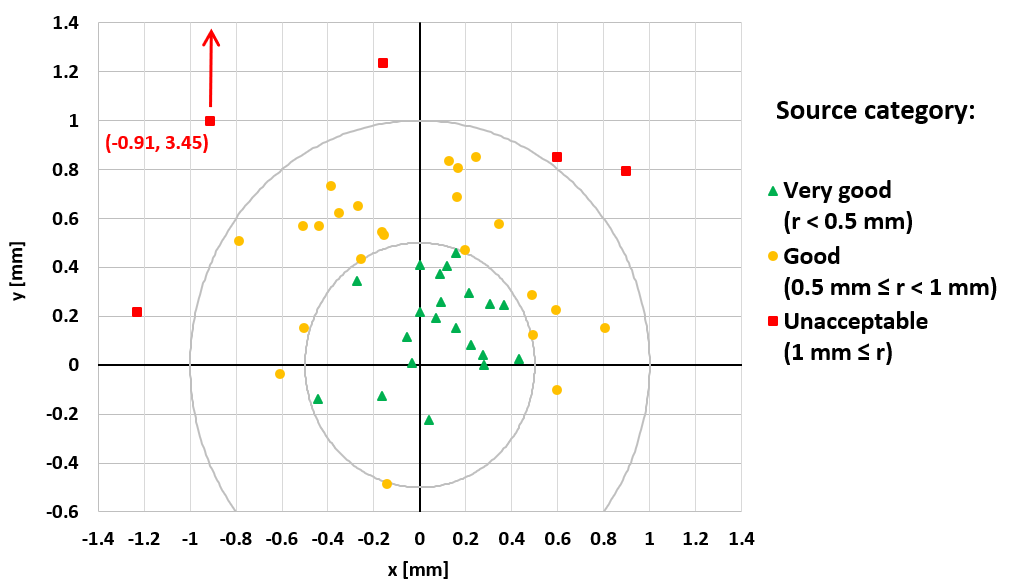}
            \caption{Position of the center of the $^{207}$Bi droplet for each of the measured sources, relative to the center of the source frame. Source quality rankings of very good (green triangle), good (yellow circle) and unacceptable (red square), are explained in section \ref{subs:categorize}.
           Uncertainties are not shown, for the purpose of clarity. Note that the position of one source fell outside the range of this plot.}
            \label{fig:distro} 
       \end{figure}

\begin{figure}
            \centering
  			\begin{subfigure}{0.36\textwidth}
  			    \centering
  			    \includegraphics[width=\textwidth]{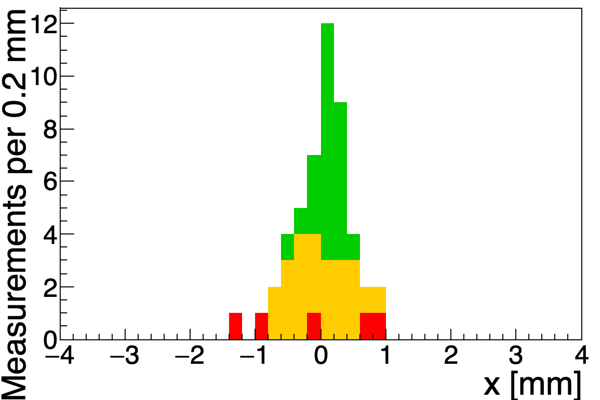}
  			    \caption{}
  			    \label{fig:x}
			\end{subfigure}
			\begin{subfigure}{0.36\textwidth}
			    \centering
  			    \includegraphics[width=\textwidth]{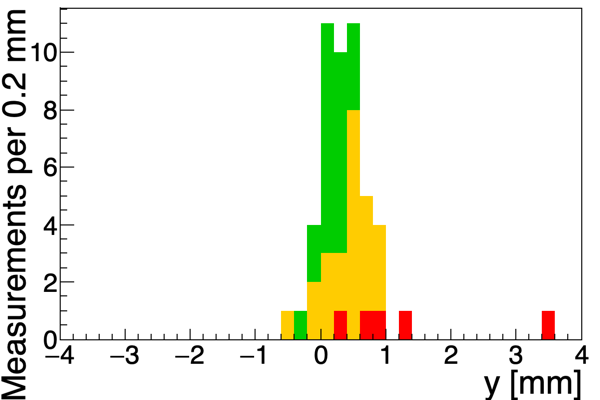}
  			    \caption{}
  			    \label{fig:y}
			\end{subfigure}
						\begin{subfigure}{0.26\textwidth}
			    \centering
  			    \includegraphics[width=\textwidth]{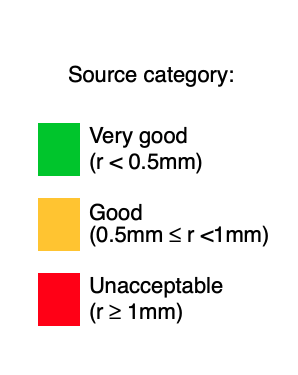}
			\end{subfigure}
			\begin{subfigure}{0.36\textwidth}
  			    \centering
  			    \includegraphics[width=\textwidth]{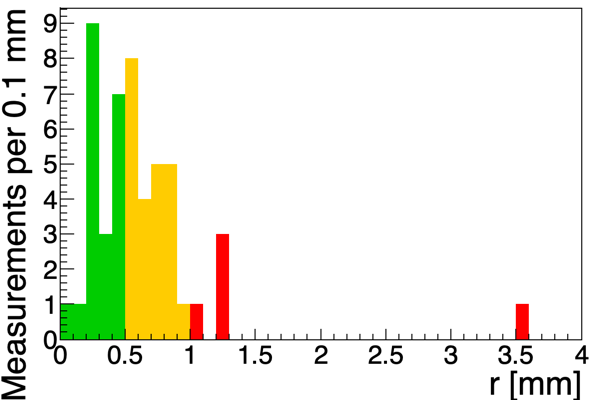}
  			    \caption{}
  			    \label{fig:r}
			\end{subfigure}
			\begin{subfigure}{0.36\textwidth}
			    \raggedright
  			    \includegraphics[width=\textwidth]{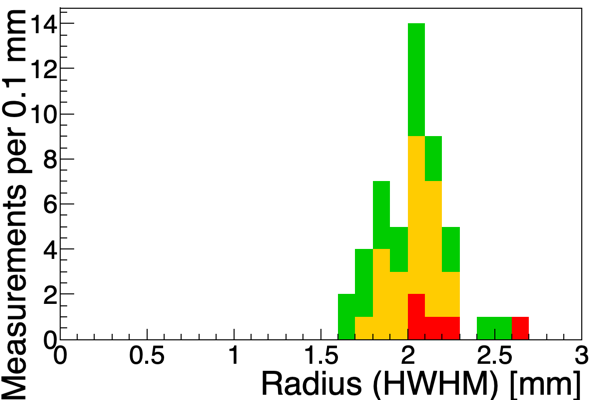}
  			    \caption{}
  			    \label{fig:hwhm}
			\end{subfigure}
			\begin{subfigure}{0.26\textwidth}
			    \centering
  			    \includegraphics[width=\textwidth]{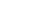}
			\end{subfigure}
			
            \caption{Summary histograms showing the positions of the droplets' centers, relative to the center of the source frame, for very good (green), good (yellow) and unacceptable (red) sources. Histograms a) and b) summarize the displacements along the short ($x$) and long ($y$) axes of the source frame respectively; c) shows the magnitudes of source droplet displacements; and d) shows the radii of the droplets. For sources where multiple measurements were taken, average values are plotted.}
            \label{fig:histos}
        \end{figure}

As previously discussed, the aim of this study was to estimate the position of the $^{207}$Bi droplet in each source relative to the center of its frame of (inner) dimensions 8\,mm~$\times$~13\,mm, where the $x$ direction corresponds to the shorter axis of the frame. This droplet position is described by the length of the vector $r=|\vec{r}|$ defined in subsection \ref{subs:datana}.
According to the magnitude of $r$ we defined three source categories:

\begin{itemize}
\item very good sources ($r <$ 0.5\,mm), represented by green triangles;
\item good sources (0.5\,mm $ \leq r <$ 1.0\,mm), represented by yellow circles;
\item unacceptable  sources ($r \geq $ 1.0\,mm), represented by red squares.
\end{itemize}
The computed values of $\vec{r}$, colored according to their category, are shown in table \ref{tab:res} for a subset of measurements, and in figure \ref{fig:distro} for all measurements (the full list is also included in tables A3 and A4 of \cite{MACKO_THESIS}). The positions of the 49 source droplets, and their radii, are summarized in figure \ref{fig:histos}.

We identified 5 sources where the center of the droplet had been deposited more than 1\,mm away from the center of the source frame center (red squares), which were rejected. The centers of the remaining 44 droplets were  within 1\,mm of the centers of the frames (green triangles or yellow circles).

\subsection{Estimation of the uncertainty on the droplet position}
\label{subs:syst}

\begin{figure}[ht]
            \centering
  			\includegraphics[width=\textwidth]{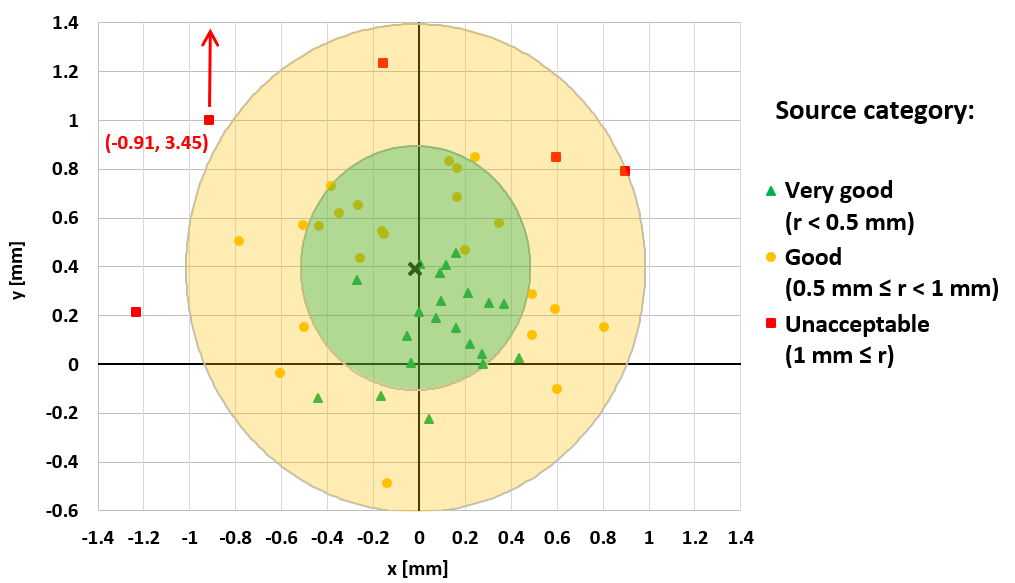}
            \caption{Position of the center of the $^{207}$Bi droplet for each of the measured sources, relative to the center of the source frame, as in figure \ref{fig:distro}. Markers denote quality rankings as in the original figure. The black cross denotes the mean position of the droplet centers at (-0.02\,mm,\,0.39 mm). Green and yellow rings denote a 0.5\,mm and 1.0\,mm offset respectively from this mean position.}
            \label{fig:meanpos} 
       \end{figure}

Before each of the measurements, the measured source was aligned to the RAP manually. In figure \ref{fig:setup_B} one can notice that in the shorter, horizontal direction ($x$) the source touched the black plastic chip holder; however, in the longer, vertical direction ($y$), the source could move freely and the human factor could introduce an extra source of uncertainty in this direction. An indication of this comes from the distribution of the measured droplet centers in \ref{fig:distro}, where it can be clearly seen that the measured $y$  positions of the centers tend to be higher than the centers of their respected source frames, as calculated from the RAP of the calibration grid (as in \ref{fig:grid}). As there is no asymmetry to the frames, and because the foils were produced with the intention of placing droplets in the center of the frames, it is reasonable to assume that the mean position of the full set of droplets should be central with respect to the frame. The offset of this mean position from our measured central point should therefore provide an estimation of the uncertainty on the measured values of the droplet position $\vec{r}$. This mean position, indicated by the black cross in figure \ref{fig:meanpos} has an offset of magnitude 0.02\,mm in the $x$ direction, and 0.39\,mm in the $y$ direction. It can be noted that, were this position to be taken as our central point, only two of our sources would be categorized as `unacceptable', with a droplet position more than 1\,mm from this new center (outside the yellow circle in figure \ref{fig:meanpos}). Both of these sources are already rejected by the original methodology.

\begin{table}[ht]
\centering
\caption{Summary of the quantities defined in figure \ref{fig:def}, for repeated measurements on an unacceptable (red), good (yellow), and very good (green) source (see section \ref{subs:categorize}). Asterisks denote a second measurement on a source, taken under the same conditions as the first.}
\begin{tabular}[c]{|c|c|cc|c|}
\hline
Source & Droplet radius &  \multicolumn{3}{c|}{Droplet displacement}  \\ \cline{2-5}
ID & HWHM &  $r_x$ &  $r_y$ & $r$  \\ 
& [mm] & [mm] & [mm] & [mm] \\ \hline
126 &   2.44 & -1.22 & 0.11 & \cellcolor[HTML] {FF0000} 1.23 \\
126* &  2.14 & -1.24 & 0.33 & \cellcolor[HTML] {FF0000} 1.28 \\
132 & 1.96 & -0.25 & 0.57 & \cellcolor[HTML]{FFD700} 0.62 \\
132* & 1.95 & -0.29 & 0.74 & \cellcolor[HTML]{FFD700} 0.79 \\
139 & 1.61 & -0.02 & -0.07 & \cellcolor[HTML]{00FF00} 0.07 \\
139* & 1.62 & -0.05 & 0.08 & \cellcolor[HTML]{00FF00} 0.10 \\ \hline
\end{tabular}
\label{tab:res}
\end{table}

As an additional check of this method of estimating the uncertainty on the droplet position, we considered the higher-statistics repeated measurements of  droplet position  and radius for three sources (126, 132 and 139), as shown in table \ref{tab:res}. Only clusters from the ROI (energy range 3--30\,keV and cluster size of one pixel) were taken into account. By comparing the droplet position vector components, $r_x$ and $r_y$, between two repeated measurements, we can (very roughly) estimate their uncertainty. The relative difference on the $r_x$ component for two different measurements of source 132 is  $0.04$\,mm. Since this value is smaller for sources 126 and 132, we considered $0.04$\,mm as the uncertainty in the  $r_x$ component, i.e. $\Delta x=0.04$\,mm. By applying the same logic for $r_y$, we obtain a value of $\Delta y=0.22$\,mm from the comparison of the two measurements for source 126.

In the $x$ direction, where source positioning was constrained by the plastic holder, the repeated-measurement method yielded a larger variation in values than was indicated by the mean position offset (0.04\,mm vs. 0.02\,mm). Conversely, in the $y$ direction, where manual alignment was required, the consideration of mean droplet position yielded a larger offset than the repeated measurements would indicate (0.39\,mm vs. 0.22\,mm). In order to be conservative, we consider that a source should be deemed unacceptable if the radius of the droplet
comes within a distance $\Delta x=0.04$\,mm or $\Delta y=0.39$\,mm of the source frame. It should be noted that, in both cases, these uncertainty estimates are small relative to the typical effective radius of the $^{207}$Bi droplet which, at around 2\,mm (see section \ref{subs:categorize}), can be understood as a very conservative estimate of the uncertainty on the source position. 

\subsection{Droplet containment within the source frames}
\label{subs:containment}

\begin{figure}[ht]
            \centering
  			\includegraphics[width=\textwidth]{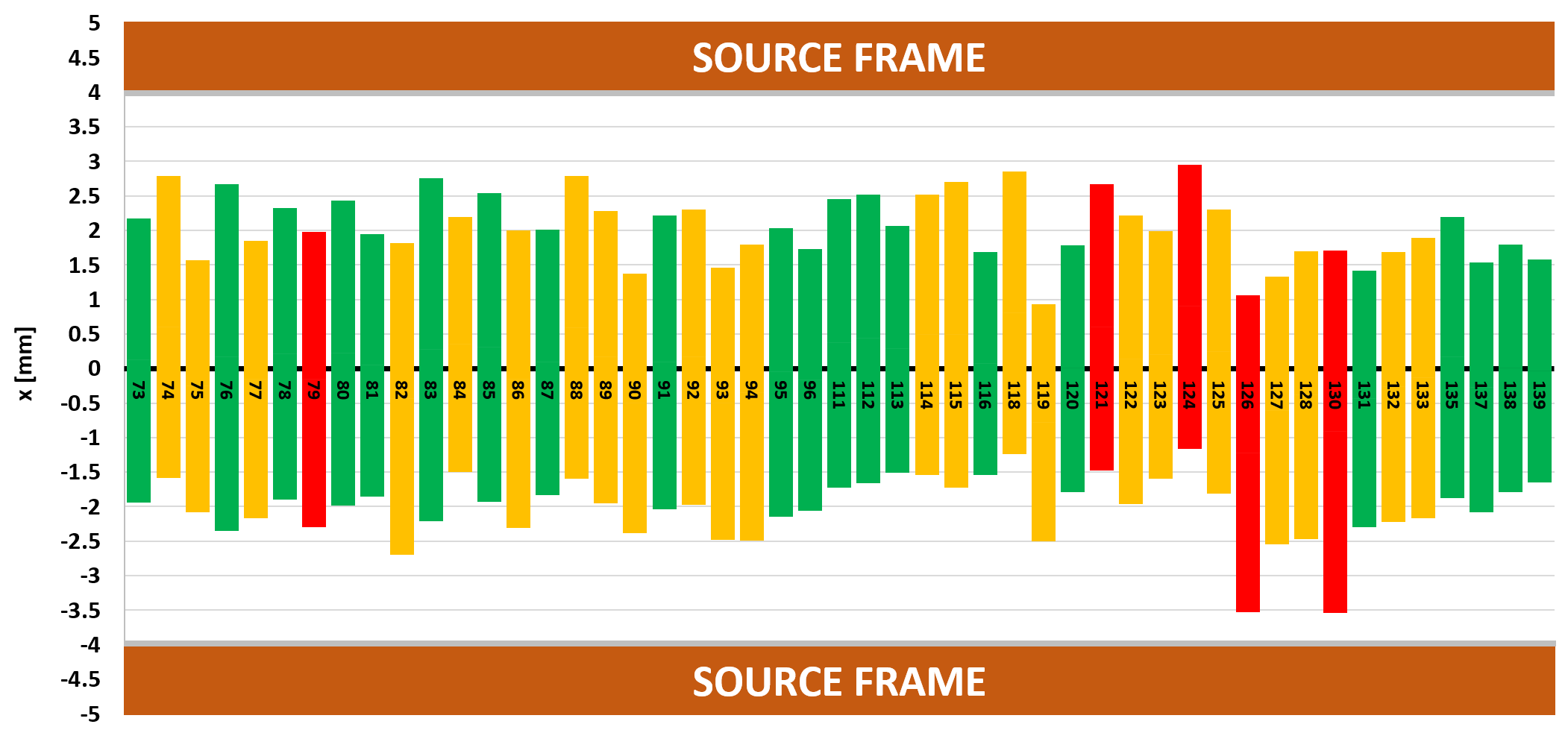}
  			\includegraphics[width=\textwidth]{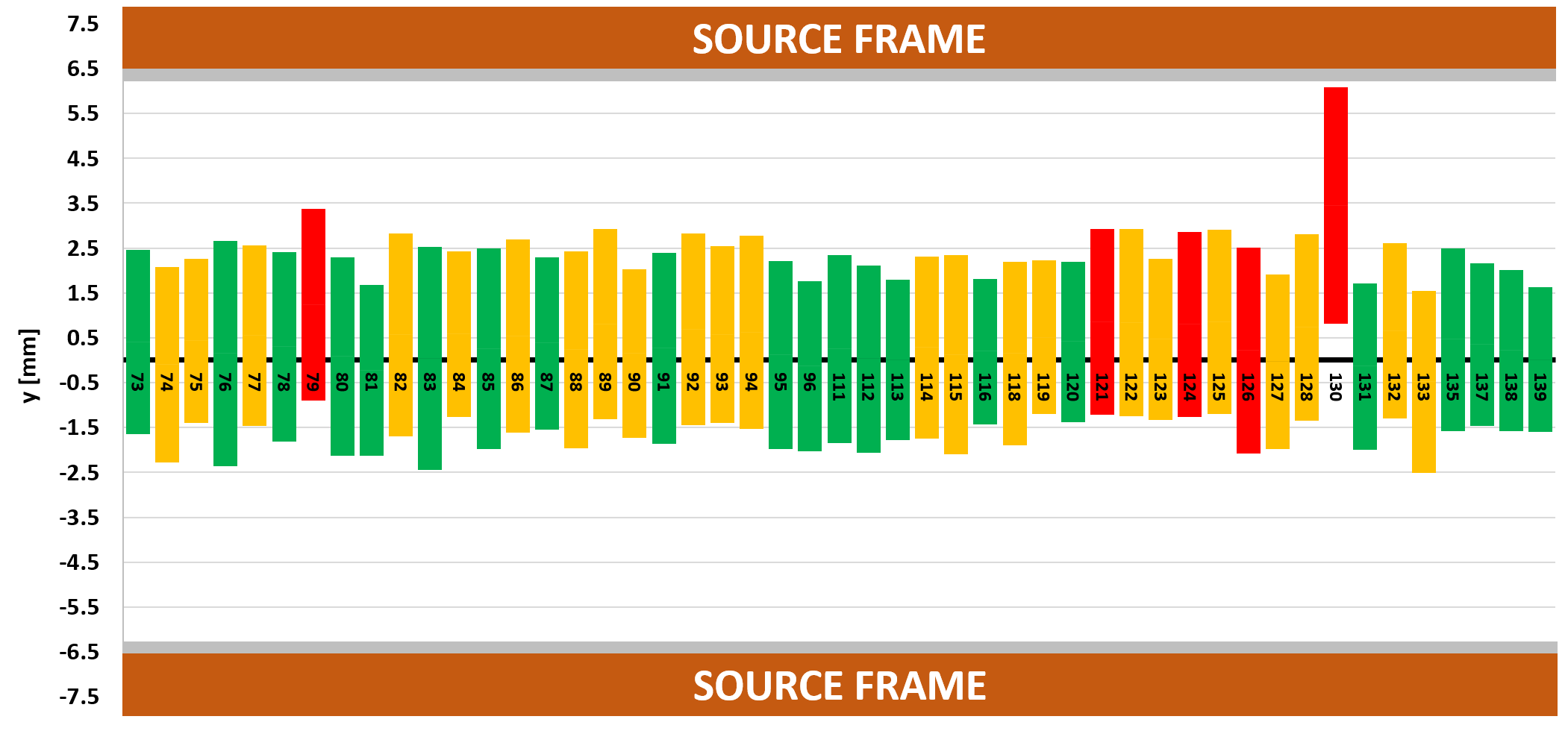}
            \caption{Projections of the $x$ (top) and $y$ (bottom) positions of the droplet in each source. The center of each bar indicates the fitted position of the droplet $r_x$ or $r_y$; the bar height is twice the HWHM; and the label indicates  source ID number. Colors indicate quality categories (see section \ref{subs:categorize}). The brown horizontal bars indicate the position of the copper frame position; the adjoining gray bars show the uncertainty on position measurements. }
            \label{fig:containment} 
       \end{figure}

As explained in section \ref{subs:datana}, one goal of this study was to ensure that sources selected for the SuperNEMO Demonstrator Module did not suffer leakage of the $^{207}$Bi droplet into the region of the copper frame. To confirm this was the case for our selected sources,  the profiles of each droplet in the $x$ and $y$ directions were plotted relative to the source frame (figure \ref{fig:containment}).

To produce the projected profiles in figure \ref{fig:containment}, each droplet is assumed to be circular, centered at its measured best-fit point, with a radius corresponding to the fitted HWHM. The projection of each droplet is shown relative to the position of the copper frame, at $\pm 4$\,mm in the $x$ direction, and  $\pm 6.5$\,mm in the $y$ direction (the frame extends beyond the plot, but only its inner dimension is relevant). Gray bars inside the source frame represent the uncertainty on position measurements in each dimension ($\Delta x=0.04$\,mm and $\Delta y=0.39$\,mm) as estimated in section \ref{subs:syst}. Sources colored green were categorized as very good, yellow as good, and red as unacceptable, based on the positions of the droplets' centers, as explained in section \ref{subs:categorize}. The mean HWHM value for all measurements, including the rejected sources and repeated measurements on individual sources, is $2.03\pm 0.21$\,mm. It can be seen that none of the selected (green and yellow) sources show any overlap between the droplet and the source frame, and furthermore, that the distance between the droplet and frame is greater than the uncertainty on the droplet's position. We therefore conclude that there is no danger of leakage under the copper frame from our selected sources.  The selected sources have a mean HWHM of $2.01\pm0.19$\,mm.

\section{Conclusions}
In this paper, we describe a method for measuring the position of $^{207}$Bi droplets in the SuperNEMO calibration sources within their frames, using Timepix detectors. We successfully applied this method to 49 $^{207}$Bi calibration sources developed for the SuperNEMO experiment, and evaluated their quality. All 49 sources  of $^{207}$Bi were proven to be eligible for calibration of the SuperNEMO Demonstrator Module, as none of them presented leaks towards the copper frame. The 42 sources whose $^{207}$Bi droplets were closest to the center of the copper source frames (out of 44 that passed our stricter quality criteria) were chosen and installed in the SuperNEMO Demonstrator Module. These measurements also allow better characterization of the source deposition distributions, ensuring that knowledge of the $^{207}$Bi source positions does not limit the energy and vertex reconstruction of the SuperNEMO Demonstrator Module.

\acknowledgments

We thank the staff of the Modane Underground Laboratory for their technical assistance in running the experiment. We acknowledge support by the grant agencies of the Czech Republic, CNRS/IN2P3 in France, RFBR in Russia (NCNIL No19-52-16002), APVV in Slovakia (Project No. 15-0576), the Science and Technology Facilities Council, part of U.K. Research and Innovation, and the NSF in the U.S. This work is supported by the Ministry of Education, Youth and Sports of the Czech Republic under the Contract Number LM2018107.

\bibliographystyle{rsc}
\bibliography{mybibfile}

\providecommand*{\mcitethebibliography}{\thebibliography}
\csname @ifundefined\endcsname{endmcitethebibliography}
{\let\endmcitethebibliography\endthebibliography}{}
\begin{mcitethebibliography}{13}
\providecommand*{\natexlab}[1]{#1}
\providecommand*{\mciteSetBstSublistMode}[1]{}
\providecommand*{\mciteSetBstMaxWidthForm}[2]{}
\providecommand*{\mciteBstWouldAddEndPuncttrue}
  {\def\EndOfBibitem{\unskip.}}
\providecommand*{\mciteBstWouldAddEndPunctfalse}
  {\let\EndOfBibitem\relax}
\providecommand*{\mciteSetBstMidEndSepPunct}[3]{}
\providecommand*{\mciteSetBstSublistLabelBeginEnd}[3]{}
\providecommand*{\EndOfBibitem}{}
\mciteSetBstSublistMode{f}
\mciteSetBstMaxWidthForm{subitem}
{(\emph{\alph{mcitesubitemcount}})}
\mciteSetBstSublistLabelBeginEnd{\mcitemaxwidthsubitemform\space}
{\relax}{\relax}

\bibitem[Arnold \emph{et~al.}(2010)Arnold\emph{et~al.}]{SUPERNEMO}
R.~Arnold \emph{et~al.}, \emph{The European Physical Journal C}, 2010,
  \textbf{70}, 927--943\relax
\mciteBstWouldAddEndPuncttrue
\mciteSetBstMidEndSepPunct{\mcitedefaultmidpunct}
{\mcitedefaultendpunct}{\mcitedefaultseppunct}\relax
\EndOfBibitem
\bibitem[Piquemal(2006)]{SNEMO-DESCR}
F.~Piquemal, \emph{Physics of Atomic Nuclei}, 2006, \textbf{69},
  2096--2100\relax
\mciteBstWouldAddEndPuncttrue
\mciteSetBstMidEndSepPunct{\mcitedefaultmidpunct}
{\mcitedefaultendpunct}{\mcitedefaultseppunct}\relax
\EndOfBibitem
\bibitem[Barabash \emph{et~al.}(2017)Barabash\emph{et~al.}]{CALORIMETER}
A.~Barabash \emph{et~al.}, \emph{Nucl. Instrum. Methods Phys. Res. A}, 2017,
  \textbf{868}, 98 -- 108\relax
\mciteBstWouldAddEndPuncttrue
\mciteSetBstMidEndSepPunct{\mcitedefaultmidpunct}
{\mcitedefaultendpunct}{\mcitedefaultseppunct}\relax
\EndOfBibitem
\bibitem[Loaiza(2017)]{FOILS}
P.~Loaiza, \emph{Journal of Physics: Conference Series}, 2017, \textbf{888},
  012086\relax
\mciteBstWouldAddEndPuncttrue
\mciteSetBstMidEndSepPunct{\mcitedefaultmidpunct}
{\mcitedefaultendpunct}{\mcitedefaultseppunct}\relax
\EndOfBibitem
\bibitem[Jeremie and Remoto(2017)]{SOURCE_PROD}
A.~Jeremie and A.~Remoto, \emph{PoS}, 2017, \textbf{ICHEP2016}, 1018\relax
\mciteBstWouldAddEndPuncttrue
\mciteSetBstMidEndSepPunct{\mcitedefaultmidpunct}
{\mcitedefaultendpunct}{\mcitedefaultseppunct}\relax
\EndOfBibitem
\bibitem[Cascella(2016)]{TRACKER}
M.~Cascella, \emph{Nucl. Instrum. Methods Phys. Res. A}, 2016, \textbf{824},
  507 -- 509\relax
\mciteBstWouldAddEndPuncttrue
\mciteSetBstMidEndSepPunct{\mcitedefaultmidpunct}
{\mcitedefaultendpunct}{\mcitedefaultseppunct}\relax
\EndOfBibitem
\bibitem[Arnold \emph{et~al.}(2005)Arnold\emph{et~al.}]{NEMO3}
R.~Arnold \emph{et~al.}, \emph{Nuclear Instruments and Methods in Physics
  Research Section A: Accelerators, Spectrometers, Detectors and Associated
  Equipment}, 2005, \textbf{536}, 79--122\relax
\mciteBstWouldAddEndPuncttrue
\mciteSetBstMidEndSepPunct{\mcitedefaultmidpunct}
{\mcitedefaultendpunct}{\mcitedefaultseppunct}\relax
\EndOfBibitem
\bibitem[Salazar and Bryant(2016)]{DEPLSYST}
R.~Salazar and J.~Bryant, \emph{PoS}, 2016, \textbf{ICHEP2016}, 808\relax
\mciteBstWouldAddEndPuncttrue
\mciteSetBstMidEndSepPunct{\mcitedefaultmidpunct}
{\mcitedefaultendpunct}{\mcitedefaultseppunct}\relax
\EndOfBibitem
\bibitem[Llopart \emph{et~al.}(2008)Llopart\emph{et~al.}]{TIMEPIX}
X.~Llopart \emph{et~al.}, \emph{Nucl. Instrum. Methods Phys. Res. A}, 2008,
  \textbf{585}, 106 -- 108\relax
\mciteBstWouldAddEndPuncttrue
\mciteSetBstMidEndSepPunct{\mcitedefaultmidpunct}
{\mcitedefaultendpunct}{\mcitedefaultseppunct}\relax
\EndOfBibitem
\bibitem[Ture\v{c}ek \emph{et~al.}(2011)Ture\v{c}ek\emph{et~al.}]{CALIB_FUNCT}
D.~Ture\v{c}ek \emph{et~al.}, \emph{2011 IEEE Nuclear Science Symposium
  Conference Record}, 2011,  1722--1725\relax
\mciteBstWouldAddEndPuncttrue
\mciteSetBstMidEndSepPunct{\mcitedefaultmidpunct}
{\mcitedefaultendpunct}{\mcitedefaultseppunct}\relax
\EndOfBibitem
\bibitem[Ture\v{c}ek \emph{et~al.}(2011)Ture\v{c}ek\emph{et~al.}]{PIXELMAN}
D.~Ture\v{c}ek \emph{et~al.}, \emph{Journal of Instrumentation}, 2011,
  \textbf{6}, C01046\relax
\mciteBstWouldAddEndPuncttrue
\mciteSetBstMidEndSepPunct{\mcitedefaultmidpunct}
{\mcitedefaultendpunct}{\mcitedefaultseppunct}\relax
\EndOfBibitem
\bibitem[Holy \emph{et~al.}(2008)Holy\emph{et~al.}]{CLUST}
T.~Holy \emph{et~al.}, \emph{Nucl. Instrum. Methods Phys. Res. A}, 2008,
  \textbf{591}, 287 -- 290\relax
\mciteBstWouldAddEndPuncttrue
\mciteSetBstMidEndSepPunct{\mcitedefaultmidpunct}
{\mcitedefaultendpunct}{\mcitedefaultseppunct}\relax
\EndOfBibitem
\bibitem[Macko(2018)]{MACKO_THESIS}
M.~Macko, \emph{Ph.D. thesis}, Universit\'e de Bordeaux; Univerzita
  Komensk\'eho (Bratislava), 2018\relax
\mciteBstWouldAddEndPuncttrue
\mciteSetBstMidEndSepPunct{\mcitedefaultmidpunct}
{\mcitedefaultendpunct}{\mcitedefaultseppunct}\relax
\EndOfBibitem
\end{mcitethebibliography}

\end{document}